\documentclass[twocolumn,useAMS,usenatbib, usegraphicx]{aastex62}
\hypersetup{linkcolor=blue,citecolor=blue}
\usepackage{aas_macros}
\usepackage{graphicx,epsfig}	
\usepackage{amsmath}	
\usepackage{amssymb}	

\begin{document}

\title[Spatially-resolved stellar populations and kinematics with KCWI: probing the assembly history of the massive early-type galaxy NGC\,1407]{Spatially-resolved stellar populations and kinematics with KCWI: probing the assembly history of the massive early-type galaxy NGC\,1407}

\correspondingauthor{Anna Ferr\'e-Mateu}
\email{aferremateu@icc.ub.edu (AFM)}

\author{Anna Ferr\'e-Mateu}
\affiliation{Institut de Ciencies del Cosmos (ICCUB), Universitat de Barcelona (IEEC-UB), E02028 Barcelona, Spain}
\affiliation{Centre for Astrophysics \& Supercomputing, Swinburne University of Technology, Hawthorn VIC 3122, Australia}

\author{Duncan A. Forbes}
\affiliation{Centre for Astrophysics \& Supercomputing, Swinburne University of Technology, Hawthorn VIC 3122, Australia}

\author{Richard M. McDermid}
\affiliation{Department of Physics and Astronomy, Macquarie University, Sydney NSW 2109, Australia}

\author{Aaron J. Romanowsky}
\affiliation{Department of Physics \& Astronomy, San Jos\'e State University, San Jose, CA 95192, USA}
\affiliation{University of California Observatories, 1156 High St., Santa Cruz, CA 95064, USA}

\author{Jean P. Brodie}
\affiliation{University of California Observatories, 1156 High St., Santa Cruz, CA 95064, USA}

\begin{abstract}
Using the newly commissioned KCWI instrument on the Keck-II telescope, we analyse the stellar kinematics and stellar populations of the well-studied massive early-type galaxy (ETG) NGC\,1407. We obtained high signal-to-noise integral-field-spectra for a central and an outer (around one effective radius towards the south-east direction) pointing with integration times of just 600s and 2400s, respectively. We confirm the presence of a kinematically distinct core also revealed by VLT/MUSE data of the central regions. While NGC\,1407 was previously found to have stellar populations characteristic of massive ETGs (with radially constant old ages and high alpha-enhancements), it was claimed to show peculiar super-solar metallicity peaks at large radius that deviated from an otherwise strong negative metallicity gradient, which is hard to reconcile within a `two-phase' formation scenario. Our outer pointing confirms the near-uniform old ages and the presence of a steep metallicity gradient, but with no evidence for anomalously high metallicity values at large galactocentric radii. We find a rising outer velocity dispersion profile and high values of the 4th-order kinematic moment -- an indicator of possible anisotropy. This coincides with the reported transition from a bottom-heavy to a Salpeter initial mass function, which may indicate that we are probing the transition region from the `in-situ' to the accreted phase.   
With short exposures, we have been able to derive robust stellar kinematics and stellar populations in NGC\,1407 to $\sim$1 effective radius. This experiment shows that future work with KCWI will enable 2D kinematics and stellar populations to be probed within the low surface brightness regions of galaxy halos in an effective way.
\end{abstract}


\section{Introduction} \label{sec:intro}
Early-type galaxies (ETGs) are a well-studied family of galaxies that tend to follow very tight relations between their key properties. Such correlations are powerful tools to understand the formation mechanism and evolution through cosmic time of ETGs and as such, must be reproduced by any valid theoretical model. A currently favored formation mechanism for ETGs is the so-called `two-phase' model (e.g \citealt{Naab2009}; \citealt{Oser2010}; \citealt{Hilz2013}). In the first phase, the core of the galaxy forms `in-situ', in a fast and violent event with extremely high star formation rates (at $z\gtrsim2$). This dissipative phase creates a compact, massive galaxy sometimes referred to as a `red nugget' (e.g. \citealt{Damjanov2009};  \citealt{Glazebrook2009}). This is then followed by an accretion phase, where the galaxy suffers random encounters with in-falling lower mass satellites. The majority of those encounters are dry mergers that place most of the newly accreted material at the periphery of the `red nugget'. This will add some stellar mass and significantly increase its size, but will leave the central properties largely unchanged. This scenario can explain many of the main phenomena seen in ETGs, such as their strong size evolution with cosmic time (e.g \citealt{Daddi2005}; \citealt{Trujillo2007}). Within this scenario, one would expect to be able to differentiate between the `in-situ' and accreted (`ex-situ') material via spatially-resolved stellar populations and kinematics, perhaps even determining the transition radius. This transition is predicted to occur at distances smaller than the effective radius (R$\mathrm{_e}$) for the most massive ETGs, owing to their very active accretion histories.(e.g \citealt{RodriguezGomez2016}).

Over the past decade, a huge step forward in understanding the formation and evolution of ETGs has been made by integral field spectroscopy (IFS) surveys (e.g. ATLAS$^{\rm 3D}$ \citealt{Cappellari2011}, CALIFA \citealt{Sanchez2012}, MaNGA \citealt{Bundy2015}, SAMI \citealt{Croom2012}). These surveys have explored the main properties of ETGs, although most of the findings are limited to the central regions of the galaxies. However, spatially-resolved studies covering the low surface brightness outer regions of ETGs are still scarce owing to limitations of instrument sensitivity and thus the large amounts of time required for such studies. 

Recently, two new IFS instruments have been installed on the Keck (Keck Cosmic Web Imager, KCWI; \citealt{Morrissey2018}) and the VLT (Multi Unit Spectroscopic Explorer, MUSE; \citealt{Bacon2010}) telescopes. Both instruments are designed for high sensitivity over long (stacked) total integration, and therefore are ideally suited to open up new avenues of study in the halos of ETGs. MUSE operates with a larger and longer wavelength coverage (480-930\,nm) than KCWI, which currently operates with bluer coverage (350-560\,nm, though with a planned future upgrade path for a red channel covering 530-1050\,nm). The bluer coverage of KCWI is, however, arguably better suited to the study of stellar populations which have more diagnostic lines at blue wavelengths, as well as taking advantage of lower background contributions from night sky emission. MUSE has the advantage of a larger field-of-view (1'$\times$1', compared to the 33"$\times$20" corresponding to the largest FOV of KCWI), though the finer spatial sampling of MUSE (0.2\arcsec\/ square spaxels vs. KCWI's $0.4\arcsec \times 1.4\arcsec$ in its largest field mode) is not necessarily an advantage for studying the extended envelopes of nearby galaxies, which can cover arcminute scales.

In this paper, we present new KCWI observations of a well-observed ETG, NGC\,1407, allowing us to test the performance of this instrument for measuring stellar kinematics and populations in different surface brightness regimes, as well as providing a comparative study with MUSE (in addition to other studies). Although too far south to be part of the ATLAS$^{\rm 3D}$ survey \citep{Cappellari2011}, NGC\,1407 was still included in the SLUGGS survey\footnote{http://sluggs.swin.edu.au} \citep{Brodie2014} as a prime example of a massive and relatively nearby ETG. Recently \hypertarget{J+18}{\citet{Johnston2018}} ( \hyperlink{J+18}{J+18} hereafter) probed NGC\,1407's central kinematics using two different MUSE datasets from the ESO archive. They totaled $\sim$4500s and the data reached out to $\sim$30" ($\sim$0.47\,R$\mathrm{_e}$). Therefore it did not reach the outer regions where \hypertarget{P+14}{\citet{Pastorello2014}} (hereafter \hyperlink{P+14}{P+14}) inferred a few locations to have super-solar metallicities using DEIMOS multi-slit spectroscopy of the Calcium II triplet (CaT). Such anomalies in the outer parts are difficult to reconcile within the `two-phase' scenario presented above, and therefore need to be understood. 

With this aim, we present here new data from the recently commissioned KCWI of both the galaxy centre and a pointing covering the anomalous region. We are thus in a position to directly compare and contrast the capabilities of KCWI with MUSE and to investigate the claimed outer metallicity peaks in NGC\,1407. We first describe our KCWI observations and data reduction in Section 3. Next we derive both stellar populations and kinematics from our new data (Section 4), and compare our values with those from the literature and the results in Section 5. 

\begin{figure*} 
\begin{center}
\includegraphics[scale=0.28]{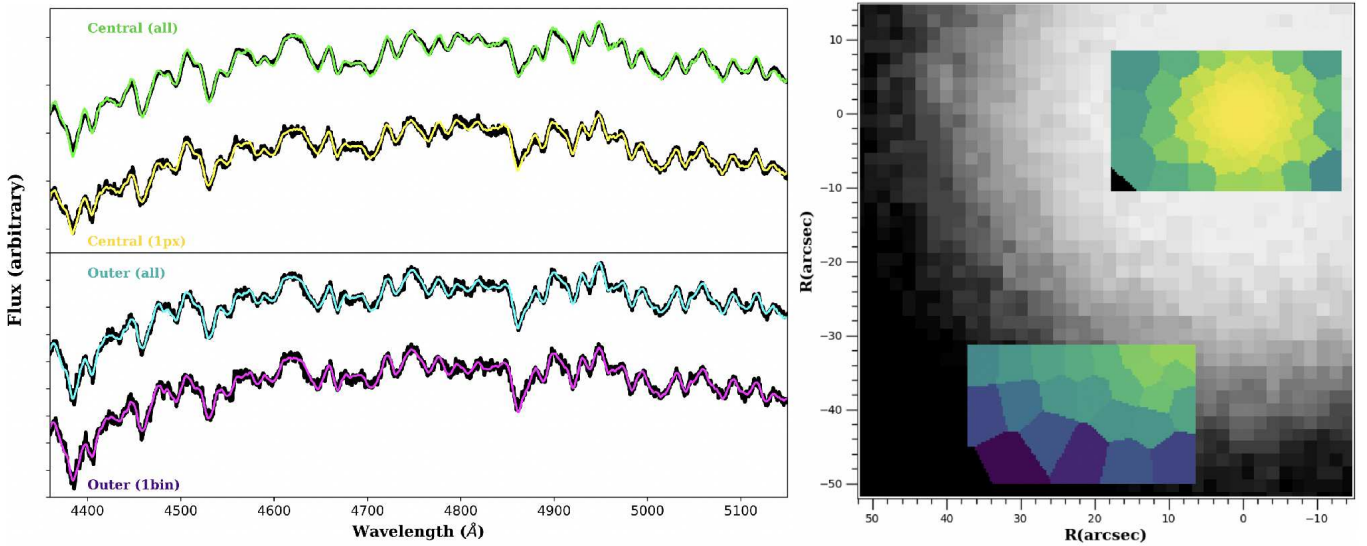}
\end{center} 
\vspace{-0.2cm}
\caption{\textbf{KCWI pointings of NGC\,1407 and the resulting spectra:} The two locations (central and outer) are shown as the tessellation map used in the analysis section, overlaid onto an DSS image of NGC\,1407. North is up and East is left. The left panels show the corresponding spectra for the central (top) and outer (bottom) pointings and the resulting fitting from  {\tt pPXF}. Two apertures are shown in each case to illustrate the quality of the data. One corresponds to collapsing the entire datacube (green for the central, cyan for the outer) whereas the others correspond to the unbinned spectra for the center-most pixel (yellow) and the outermost bin of the outer pointing (magenta).}
\end{figure*}

\section{NGC 1407}
Located at the centre of the massive Eridanus group (\citealt{Gould1993}), NGC\,1407 is the brightest group galaxy, classified as an E0. Its stellar mass is log(M$_{\ast}$/M$_\odot$)\,$\simeq$\,11.60 \citep{Forbes2017}) and its total halo mass is log(M$\mathrm{_h}$/M$_\odot$)\,$\simeq$\,13.2 \citep{Wasserman2018}. Here we take the distance to NGC\,1407 (26.8 Mpc) and its effective radius (R$\mathrm{_e}$ = 63.4") from \hyperlink{P+14}{P+14} to enable direct comparison.

The galaxy has been extensively studied both with classical long-slit spectroscopy and with IFS. It has a recession velocity of $v_{r}$=1779 km\,s$^\mathrm{-1}$ and shows a well defined kinematically distinct core (KDC). This was first reported in \citet{Spolaor2008} using long-slit spectroscopy, although the data were not sufficient to determine if it was a true feature. Since then, its kinematics in the central parts have been followed up by \citet{Proctor2009}; \citet{Rusli2013}; \citet{Arnold2014} and \citet{Foster2016}. More recently, a new attempt to measure the size ($r\sim$ 5 arcsec or 0.6\,kpc) and amplitude of the KDC was performed with MUSE by \hyperlink{J+18}{(J+18)}. While they quote an amplitude of $\sim$\,10\,km\,s$^{-1}$, this is lower than that \citet{Rusli2013}. Because the MUSE data was more coarse and seeing-limited, no conclusive results were reached. 

Using long-slit data along the major axis, \hypertarget{S+08b}{\citet{Spolaor2008a}} (\hyperlink{S+08b}{S+08b} hereafter) showed that NGC\,1407 is very old with a flat age gradient but a steep metallicity gradient. Also using long-slit spectroscopy along the major axis, but of much higher signal-to-noise (S/N), \citet{Conroy2017} and \hypertarget{vD+17}{\citet{vanDokkum2017}} (\hyperlink{vD+17}{vD+17} hereafter) studied its stellar populations and initial mass function (IMF). Their results are consistent with \hyperlink{S+08b}{S+08b} and in addition, they also found evidence for a very bottom-heavy IMF within the central $\sim$0.3\,R$\mathrm{_e}$. Although the main focus in \hyperlink{J+18}{J+18} was on the kinematic features of NGC\,1407, they also performed a tentative stellar population analysis with similar results. In all three works, the authors found high values of [$\alpha$/Fe] at all radii, indicative of fast formation timescales in ETGs \citep{Thomas2005}.

Given the above, NGC\,1407 seems to be a rather typical massive ETG. However, an IFS study of the stellar populations in the outer regions of NGC\,1407 has been carried out only by the SLUGGS survey using the pseudo-IFS method called SKiMS. Because it uses the CaT absorption lines, stellar metallicities can be inferred but there is no information on the stellar ages. Intriguingly, \hyperlink{P+14}{P+14} claimed that the smooth radial decrease in metallicity showed several locations of super-solar metallicity ([Z/H]$>$0.8\,dex) around 45" ($\sim$0.8\,R$\mathrm{_e}$) south-east of the galaxy centre. If correct, they could represent dynamically unmixed stars from a previous major merger. It is therefore important to confirm or refute these claimed off-centre metallicity peaks with new IFS data. One aim of this paper is to further investigate these anomalous metallicity peaks, while providing the first 2D maps of a massive ETG out to $\sim$1\,R$\mathrm{_e}$ with KCWI.

\section{Observations and data reduction}
Observations were carried during the night of 2017 September 19th as part of the first public run of the newly commissioned instrument KCWI on the Keck-II telescope. NGC\,1407 was a filler object in a larger program, aimed at testing the capability of obtaining good IFS data in the outer regions of an ETG with a modest amount of time in dark time conditions. The weather conditions were irregular with clouds and a typical seeing of 1.5". 

We employed the large slicer (33" $\times$ 20") with a position angle of 0\,deg. It was used with the blue BM grating centered at $\lambda$=4800\AA, which gives a usable wavelength range of 4300--5200\,\AA. While this is an excellent regime to perform studies of stellar population age and metallicity, the Mg$\mathrm{_b}$ index is not available as it falls at the limit of our spectra, and therefore we cannot derive any alpha-element information for use the Mg$\mathrm{_b}$ absorption line to derive the total metallicities ([Z/H]). This configuration delivers a nominal spectral resolution of $\sim$2.5\,\AA \,($\sigma \sim$ 64 km\,s$^\mathrm{-1}$) at 5000\AA. However, measuring the spectral resolution from the arc lamps gives a virtually constant higher resolution of $\sim$2.0\,\AA \,($\sigma \sim$ 55 km\,s$^\mathrm{-1}$; R$\sim$2300) throughout the wavelength range. This is marginally better than the MUSE resolution of $\sim$2.7\,\AA \,($\sigma \sim$ 69 km\,s$^\mathrm{-1}$; at 5000\AA; \hyperlink{J+18}{J+18}). 

\begin{figure*} 
\begin{center}
\includegraphics[scale=0.32]{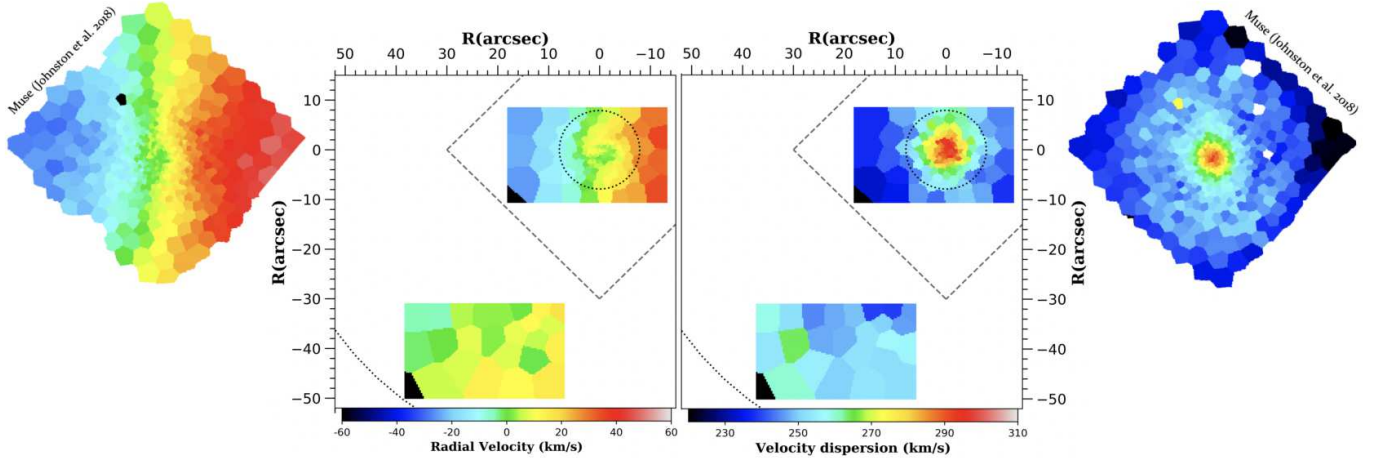}
\end{center} 
\vspace{-0.2cm}
\caption{\textbf{2D stellar kinematic maps of NGC\,1407:} The KCWI low-order kinematics of NGC\,1407 are shown, with the rotation velocity in the left panel and the velocity dispersion in the right panel. The small stamps on the sides (courtesy of E. Johnston) show the MUSE kinematic results from \hyperlink{J+18}{J+18} corresponding to the dashed diamond area in our panels, using the same color-scale. The dotted circles correspond to R$\mathrm{_e}$/8 (smaller) and 1\,R$\mathrm{_e}$ (larger). The recovered kinematics in the central region are almost identical, including the presence of a KDC.} 
\end{figure*}

We integrated for 600s for the central pointing and obtained 4 exposures of 600s (2400s total) for the outer pointing at $\Delta_\mathrm{RA}$=20", $\Delta_\mathrm{Dec}$=$-$40", i.e., to the south-east, as shown in Figure 1. With this we reach roughly 1\,R$\mathrm{_e}$, which corresponds to a surface brightness of $\mathrm{\mu_B}\sim$\,22\,mag per sq. arcsec. No nod-and-shuffle was applied, but we observed instead a nearby patch of sky for 600s to later perform sky subtraction, and a standard star was observed for flux calibration purposes. 

The data were reduced using the KCWI pipeline KDERP\footnote{github.com/Keck-DataReductionPipelines/KcwiDRP}, which performs a full data reduction, delivering flux calibrated datacubes. For the sky subtraction we tested both options of the pipeline. One subtracts the pure sky frames from the science targets, whereas the second option, available in a later updated version of the pipeline, applies a 1--sigma clipping to remove light from non-sky photons. However, this second method is limited by needing enough sky pixels for the clipping procedure, which is not the case in our central pointing. We therefore choose the outputs from applying the initial technique of dedicated sky frames during the observations.

\section{Analysis}
Once the science datacubes are complete we perform a Voronoi tessellation (e.g \citealt{Cappellari2003}) to spatially bin the spectra and optimize the S/N in each bin to $\sim$30. We note that we have also tested lower quality thresholds, down to S/N$\sim$10. While the trends described below are reproduced in these lower S/N binnings, finer trends are diluted by the higher uncertainties in some of the bins.

\begin{figure} 
\begin{center}
\includegraphics[scale=0.58]{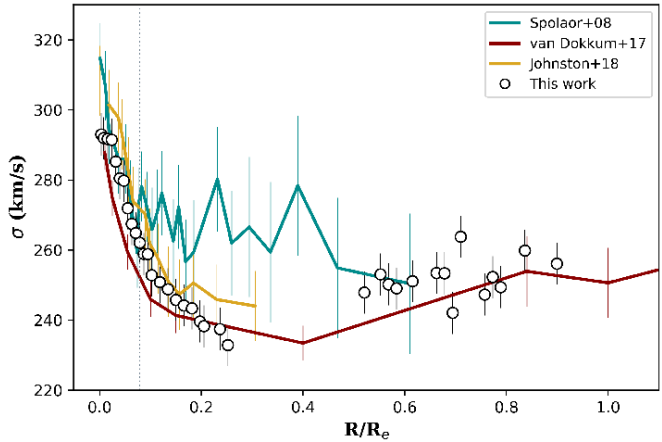}
\end{center} 
\vspace{-0.2cm}
\caption{\textbf{Radial velocity dispersion profile of NGC\,1407:} The 1D radial profile for the measured velocity dispersion from different literature works is shown. We find good agreement with literature studies, showing an increase outside of $\sim$0.3--0.4R$\mathrm{_e}$. The dotted vertical line marks the size of the KDC from \hyperlink{J+18}{J+18}.}
\end{figure}

These high S/N tessellated maps are fed into {\tt pPXF} (Penalized Pixel Fitting; \citealt{Cappellari2004}) in order to obtain the stellar kinematics (radial velocity, velocity dispersion, and the higher-order moments $h_3$ and $h_4$), together with the stellar populations (age, metallicity, and alpha-enhancement). For both the kinematics and the stellar populations, we employ the E-MILES SSP library \citep{Vazdekis2016}, considering templates that range from metallicities of $-$2.42 to $+$0.40\, dex and that cover ages from 0.03 to 14\,Gyr.

\begin{figure*} 
\begin{center}
\includegraphics[scale=0.29]{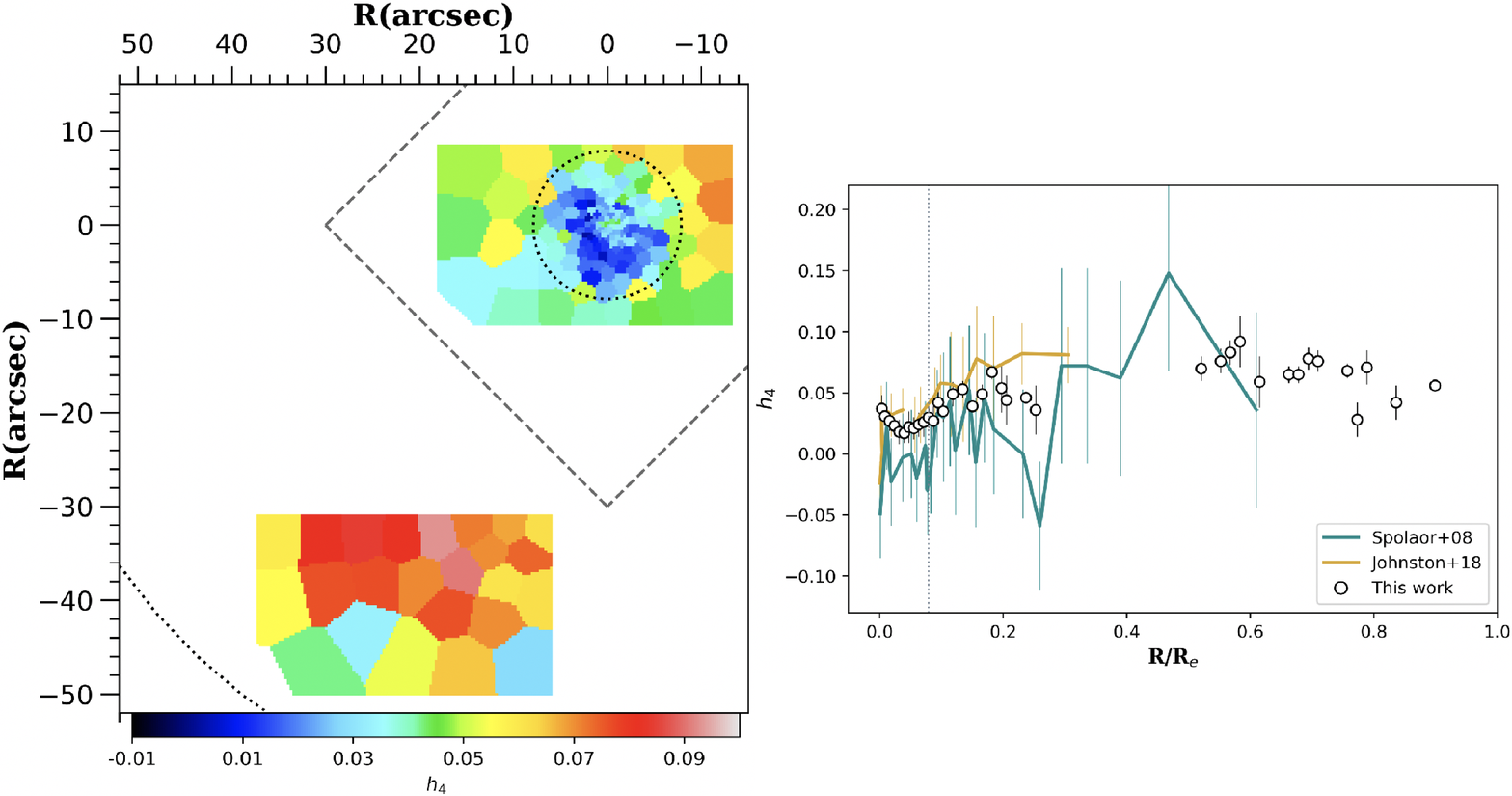}
\end{center} 
\vspace{-0.2cm}
\caption{\textbf{2D map and radial gradients of the fourth-order Gauss-Hermite moment in NGC\,1407:} \textit{Left panel:} there is a strong positive increase of the values of $h_{4}$ from the central to $\sim$0.6R$\mathrm{_e}$, when it then seems to decrease. \textit{Right panel:} our data follow closely the values of \hyperlink{J+18}{J+18}. The \citet{Spolaor2008} values show a similar trend albeit in general lower. The dashed vertical line marks the size of the KDC from \hyperlink{J+18}{J+18}.}
\end{figure*}

We have also investigated the results from using different approaches. For example, we tested different alpha-element models, as NGC\,1407 is known to have a high [Mg/Fe]$\sim\,+0.35$~dex ratio (e.g. \hyperlink{vD+17}{vD+17}; \hyperlink{J+18}{J+18}). We have also used a varying IMF, as NGC\,1407, like most massive ETGs (e.g. \citealt{Martin-Navarro2015}), shows a strong radial IMF gradient, going from a very bottom-heavy IMF in the central parts to a universal Kroupa-like IMF at around 0.4~R$\mathrm{_e}$ \hyperlink{vD+17}{(vD+17)}. This is relevant as the use of a non-universal IMF has been shown to affect the stellar populations analysis \citep{Ferre-Mateu2013}. We note, however, that such tests show only a small impact on the mean mass-weighted ages and leave the total metallicity values largely unchanged. Therefore, due to the lack of both alpha enhanced and IMF sensitive features in our spectral range, and to allow for direct comparison with the literature, we present hereafter the results utilizing the Basti Base SSP models and a Kroupa IMF.

\subsection{Kinematics}
We present briefly our kinematic results to show how they compare with published data. However, we encourage the reader to refer to \hyperlink{J+18}{J+18} for a more comprehensive study of the stellar kinematics of NGC\,1407, in particular their discussion on the nature of NGC\,1407's KDC. Overall we find an excellent agreement in the kinematic features of the central parts of NGC\,1407 with those in \hyperlink{J+18}{J+18}, as seen by the 2D kinematic maps in Figure 2. The left panel shows a slow rotational velocity of $\sim$40 km\,s$^\mathrm{-1}$ and the small KDC is clearly seen. We note that the amplitude of the KDC we observe is similar to the one in \hyperlink{J+18}{J+18} ($\leq$10\,km\,s$^\mathrm{-1}$), as opposed to the larger value of 40\,km\,s$^\mathrm{-1}$ from \citet{Rusli2013}. However, we could be suffering from the same effects as \hyperlink{J+18}{J+18} owing that both KCWI and MUSE have a similar resolution, which is coarser than the one of SINFONI \citep{Rusli2013}, and that they are both seeing-limited. In fact, there is a peculiar small feature in the KDC that after inspection in the different S/N bins disappears as we increase the S/N threshold. This indicates that it is most likely a noise feature rather than a real feature. Nonetheless, it has no impact in our results, as shown in the following section. 

The right panel of Figure 2 presents a peaked velocity dispersion of $\sim$300 km\,s$^\mathrm{-1}$ at the centre that decreases to 240\,km\,s$^\mathrm{-1}$ at $\sim$0.3--0.4\,R$\mathrm{_e}$. However, beyond that point the $\sigma$ profile begins to rise, reaching values of 260\,km\,s$^\mathrm{-1}$ at 1\,R$\mathrm{_e}$. This rising profile is better seen in Figure 3. Our declining velocity dispersion part is well-matched to the major axis profiles of \citet{Spolaor2008} and \hyperlink{vD+17}{vD+17}, with the latter also showing the rising $\sigma$ beyond 0.4\,R$\mathrm{_e}$. 

Such outwards rising profiles can only be revealed when reaching large galactocentric distances, as shown by the MASSIVE survey \citep{Ma2014}. Although NGC\,1407 was not included in that survey, it would lie at the lower mass limit (log(M$_{\ast}$/M$_\odot$) $>$ 11.6) of their sample. \citet{Veale2018} found that around $\sim$15$\%$ of the MASSIVE galaxies presented this rising profile (\textbf{see also \citealt{Ene2019}}). However, \citet{Veale2018} also found that massive galaxies with such a rising $\sigma$ profiles also have increasing values of the 4th Gauss-Hermite coefficient ($h_{4}$) with radius. As shown in Figure 4, the 2D map of NGC\,1407's $h_{4}$ and its radial gradients show this increase up to $\sim$0.6R$\mathrm{_e}$ (also reported in \hyperlink{J+18}{J+18}). However, it then hints at a possible decrease or flattening of the $h_{4}$ values, which was also found in the pseudo-IFS maps of \citealt{Arnold2014} (although this latter found lower values of $h_{4}$ on average). We discuss the implications of these kinematics profiles together with the outcome from the stellar populations in Section 5.

\begin{figure*} 
\begin{center}
\includegraphics[scale=0.50]{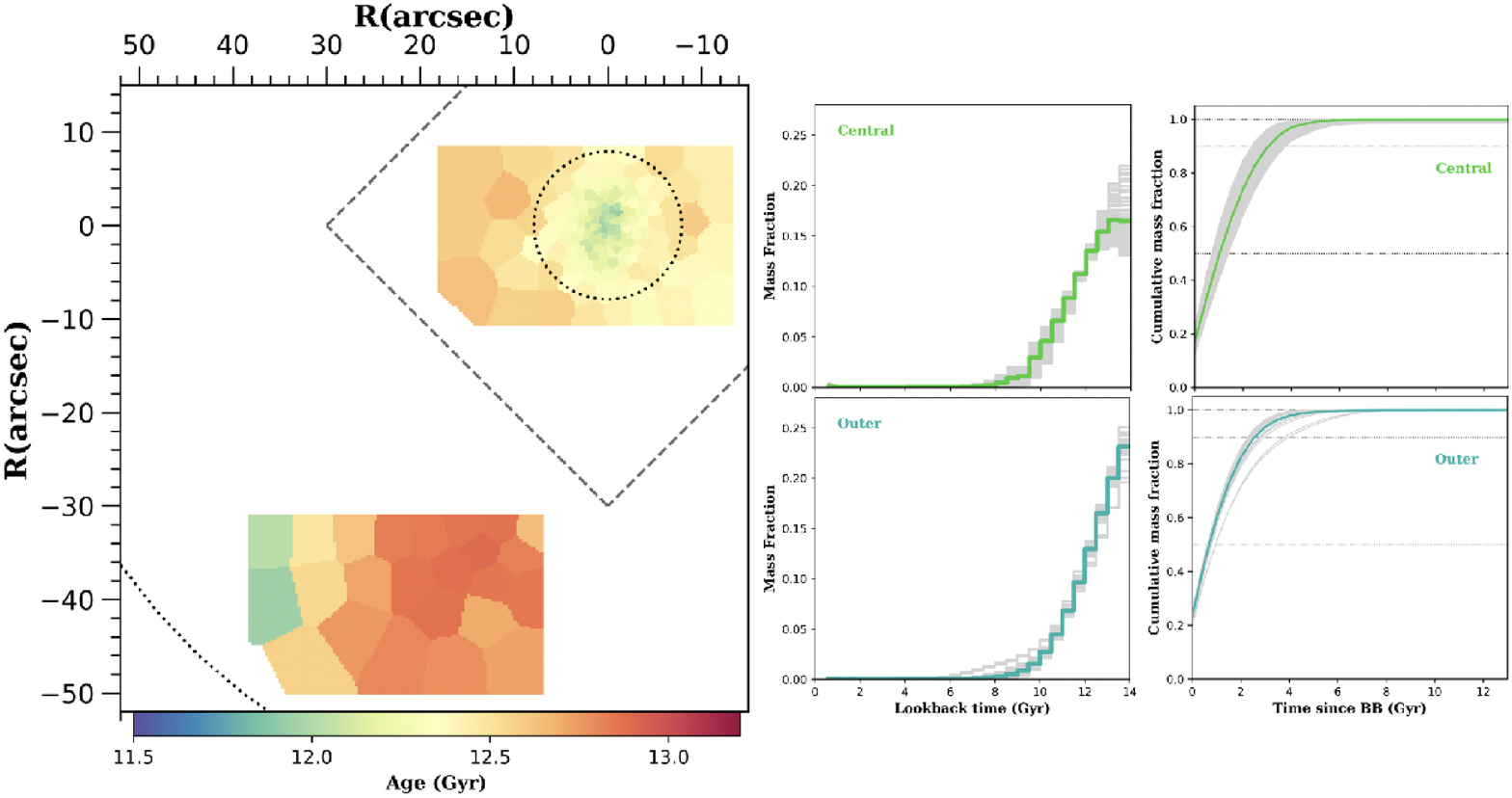}
\end{center} 
\vspace{-0.2cm}
\caption{\textbf{2D map of the stellar ages in NGC\,1407 and its SFH:} \textit{Left panel:} The 2D map shows that NGC\,1407 is uniformly old, with a hint of slightly younger ages in the center. Note that there is a steep IMF out to 0.3R$\mathrm{_e}$, which corresponds to $\sim$30". \textit{Right panels:} the SFHs and cumulative mass for each bin (grey) for the central (top) and outer (bottom) pointings. The solid green and cyan lines represent the averaged SFH and cumulative mass of the central and the outer pointings, respectively.  In the cumulative mass panels we mark  with dotted lines where the galaxy has built up 50, 90, and 100$\%$ of its stellar mass.}
\end{figure*}

\subsection{Stellar Populations}
We also use {\tt pPXF} to obtain the main stellar population parameters (mean age and total mean metallicity [Z/H]) and the star formation history (SFH) of NGC\,1407. In the following we present the results of using {\tt pPXF} with regularization. The regularization value was derived following the procedure described in \citet{McDermid2015} and \citet{Cappellari2017}. Using such regularization provides a trade-off between a smoother SFH and a good fit that is still consistent with the observations. Note that a fit without regularization in the case of a massive ETG like NGC\,1407 would provide a result that is more similar to the single stellar population approach (i.e. using a line index analysis) presented in the literature. As mentioned above, due to the lack of a reliable Mg$\mathrm{_b}$ index value, we are not able to measure our own [Z/H] from the line indices, hence compare it directly to the literature values. However, we do perform a line-index analysis to derive the iron metallicity [Fe/H] that will be later discussed in Section 5, using the pair of indices H$\mathrm{_\beta}$--Fe5015. 

The derived SFHs for all bins (see Figure 5, right panels) are fairly similar in terms of the timescales: they all show that the galaxy created stars very early on with fast star-forming rates, being assembled in a timescale of less than 3\,Gyr. This fits well into the general scheme of formation for massive ETGs \citep{Thomas2005}, where such short formation timescales are translated into elevated alpha-abundances (e.g \citealt{Thomas2005}; \citealt{delaRosa2011}, \citealt{McDermid2015}). Indeed, NGC\,1407's [$\alpha$/Fe] was previously reported to be constantly high at all radii ([$\alpha$/Fe]$\sim\,+$0.35\,dex; e.g \hyperlink{vD+17}{vD+17}; \hyperlink{J+18}{J+18}). NGC\,1407 is consistently old on average ($\sim$12--13\,Gyr) at all the radii covered by our two pointings, which is in agreement with other literature studies (e.g \hyperlink{S+08b}{S+08b}; \hyperlink{vD+17}{vD+17}). The 2D stellar age map shows a hint of slightly younger central ages (similarly to \hyperlink{J+18}{J+18}). Although the difference in age is compatible with the associated error, the region roughly corresponds to the coverage of the KDC, indicating that both phenomena might be related and that those young ages are a real feature and not systematics of the SSP models. Such slightly younger ages could hint towards a more recent event of star formation, such the occurrence of a gas-rich major merger. 

We find that NGC\,1407 shows a strong metallicity gradient towards the outer pointing. In Figure\,6 we show the 2D metallicity map of NGC\,1407. It has a super-solar central metallicity of [$Z$/H] $\sim$ $+$0.27\,dex that decreases to $\sim -0.10$~dex at $\sim$1R$\mathrm{_e}$. We compare our metallicity gradient of NGC\,1407 with those in the literature (see Figure\,6, right panel). This includes the long-slit gradients of \hyperlink{S+08b}{S+08b} and \hyperlink{vD+17}{vD+17}, both of whom oriented their long-slit along the major axis of the galaxy at $\sim$40$^\circ$ (our outer pointing lies closer to the minor axis). We also show the metallicity profiles from \hyperlink{J+18}{J+18} (adapted from their fig. 5) and from the 2D pseudo-IFS of \hyperlink{P+14}{P+14} (adapted from their fig. 13). We measure a metallicity log-gradient of $\Delta$[$Z$/H]$\sim -$0.20\,dex per dex like \hyperlink{vD+17}{vD+17}. \hyperlink{S+08b}{S+08b} and \hyperlink{J+18}{J+18} found a steeper gradient of $\Delta$[$Z$/H]$\sim -$0.35 dex per dex, albeit over a more limited radial range.

\begin{figure*} 
\begin{center}
\includegraphics[scale=0.29]{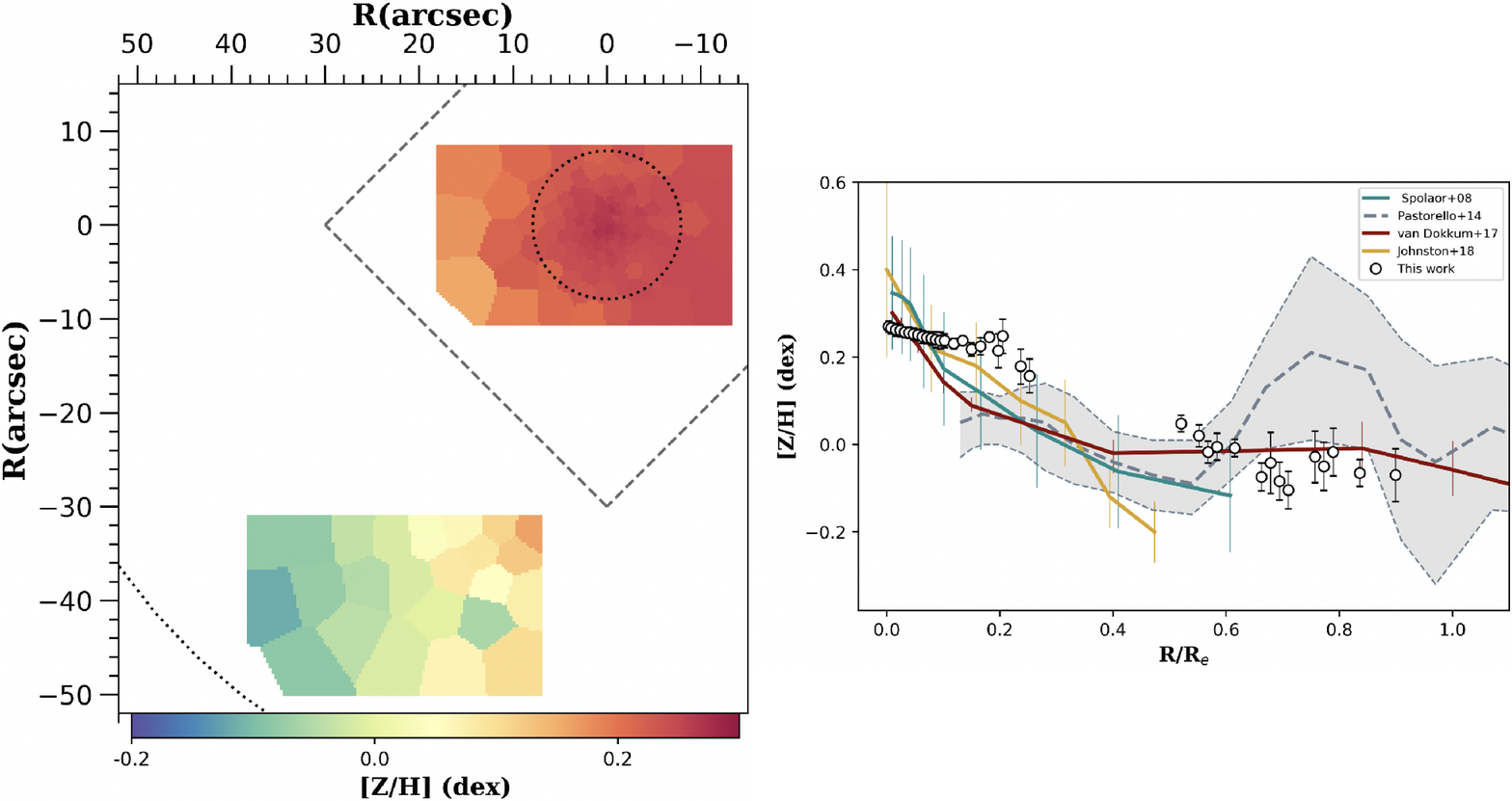}
\end{center} 
\vspace{-0.2cm}
\caption{\textbf{2D map and radial profile of the metallicity in NGC\,1407:} \textit{Left Panel:} the central and outer pointings reveal a strong radial negative metallicity gradient, with no evidence for anomalously high metallicities in the outer pointing. \textit{Right panel:} metallicities from various literature studies are shown along with our KCWI measurements in two pointings (open circles).}
\end{figure*}

 We note that the central metallicities of the literature studies tend to be higher than ours and thus show steeper metallicity gradients. There are two main contributors to this effect. From one side, the use of the line index technique tends to provide slightly higher metallicities (uses one single SSP) as opposed to the full-spectral-fitting technique (which provides a combination of SSPs that better fit the spectra; see e.g \citealt{Ferre-Mateu2014}). From the other side, the use of a regularization parameter creates a smoother (and thus less SSP-like) SFH. These effects will be most prominent at the center-most parts, where the galaxy behaves more like a single-burst dissipative event, whereas the outer regions are expected to have a more extended SFH as the result of the accretion phase. Therefore, a non regularized fit or the use of line indices can provide a slightly higher metallicity in the center parts, thus producing steeper gradients. In fact, our non-regularized test recover the higher metallicities shown in the central parts of the galaxy. However, the regularization parameter employed here is the one that provides the best balance between a good fit that is consistent within the errors and a smoother SFH \citep{McDermid2015}.

Furthermore, it is worth emphasizing that regardless of which SSP models we employ (scaled solar, with varying alphas, different IMF, etc.) or regularization value, we find that the metallicities in the outer pointing are \textit{always} less than solar, and thus the overall results and conclusions remain unchanged. That is, we do not find the super-solar metallicities ([$Z$/H]$\sim\,+$0.8$\pm$0.15\,dex) inferred by \hyperlink{P+14}{P+14} around 0.8~R$_e$. The main reason for this discrepancy can be explained by the methodology employed. While CaT is a good metallicity indicator, it is highly dependent on the assumed IMF and additionally, it relies on an empirical transformation from the measured CaT index to an inferred metallicity. In the case of the high CaT values reported by \hyperlink{P+14}{P+14} they are extrapolated well above the empirical relation. Thus we conclude that the anomalous metallicities reported by \hyperlink{P+14}{P+14} are not most likely over-estimated, and therefore NGC\,1407's stellar population properties are similar to other ETGs of similar mass.

\section{Discussion}
We have explored the stellar populations and kinematics of NGC\,1407 covering two different regions of interest: the center-most and an outer region in the south-east, claimed to have atypically high metallicities that are hard to reconcile within a `two-phase' formation paradigm. However, the overall results show that NGC\,1407 is rather a normal ETG. Nonetheless, much more information can be revealed from these new data, completing the puzzle of the formation and assembly of NGC\,1407. 

NGC\,1407's relevant kinematic features can be closely linked to its past formation events. The presence of a KDC in NGC\,1407 suggests that this galaxy may have experienced either a major merger at some point in the past (e.g. \citealt{Hernquist1991}; \citealt{Jesseit2007}; \citealt{Bois2011}) or a series of colissionless events (e.g. \citealt{Holley-Bockelmann2000}). \citealt{McDermid2006} showed that for the SAURON \citep{Bacon2001} galaxies with a KDC, those with large KDCs (diameters larger than 1\,kpc) were all typically old ($>$10\,Gyr) and slow rotators. \hyperlink{J+18}{J+18} reported the radius of the KDC in NGC\,1407 to be around 5 arcseconds ($\sim$0.6\,kpc), therefore NGC\,1407 would fall into this category as it is a slow rotator. In addition, although NGC\,1407 is on average old ($\sim$13\,Gyr) the region of the KDC coincides with the physical region in the center revealing slightly younger ages (see Figure 5). This indicates that such young ages are real and linked to the KDC and are not systematics related to the SSP models. The most plausible explanation is the occurrence of a gas--rich major merger at slightly later times than the formation of the host galaxy. This can deliver younger stars and/or any gas left in the core of the host galaxy, producing further star formation and therefore revealing the slightly younger ages seen in the center. 
Interestingly, \citet{Forbes2018} presented a simulated ETG from {\tt Magneticum} \citep{Remus2017} that was created to reproduce a NGC\,1407-like ETG precisely. As shown in their figure 5, the model galaxy is a mixture of a major, and some minor, merger events at later times. In the simulation, the major merger event took place at $z\sim$1.1 (around 8 Gyr ago). Since our SFHs seem to be all complete around 9\,Gyr ago, this indicates that such a major merger event must have occurred earlier for the real NGC\,1407, but it is compatible with the scenario of an early major merger event indicated by the younger central ages.

The other kinematics orders of NGC\,1407 also provide key information about the formation of this massive ETG. For example, the $h_{4}$ value of NGC\,1407 is found to be positive and increasing up to $\sim$0.6R$\mathrm{_e}$. A decrease is then hinted, which was already shown in \citealt{Arnold2014}. An increase of the $h_{4}$ is associated with radial velocity anisotropy, whereas a decrease of it is related to a tangential anisotropy (e.g \citealt{Gerhard1998}; \citealt{Dekel2005}; \citealt{Thomas2007}). Previously, \citet{Pota2015} and \citet{Wasserman2018} showed that the orbits of NGC\,1407's globular clusters were radial for the red subpopulation and tangential for the blue one. Because red globular clusters are thought to track the stellar light of the galaxy (e.g. \citealt{Brodie2006}; \citealt{Peng2006}; \citealt{Forbes2018}), our results from the stellar kinematics seem to reproduce the same behavior as the red globular clusters of NGC\,1407. Typically, red globular clusters are concentrated in the central parts of the galaxies, being mostly associated with the first phase of formation (e.g. \citealt{Brodie2006}; \citealt{Beasley2018}). Blue globular clusters are instead found at all radii and are mostly associated with the accretion phase. 
Furthermore, \citet{Greene2019} investigated structural parameters such as $h_4$ that could correlate directly with the amount of `ex-situ' (accreted) material. They found that galaxies from the MASSIVE survey with high [$\alpha$/Fe] and low [Fe/H] in the outskirts tend to have positive $h_{4}$ values. Using a line index analysis we find a gradient of $\Delta$[Fe/H] = $-0.27$ dex per dex in NGC\,1407, which is fully consistent with the quoted mean value of $\Delta$[Fe/H]\,=\,$-0.26$\,$\pm$\,0.04 dex per dex for ETGs of NGC\,1407's stellar mass and velocity dispersion in \citet{Greene2019}. Therefore, NGC\,1407, with an average positive $h_4\sim$0.07, high [$\alpha$/Fe]$\sim$\,0.35\,dex and low [Fe/H]$\sim\--$0.47\,dex at the outermost regions of our data, shows the same features of the MASSIVE galaxies in \citet{Greene2019}. As mentioned above, positive values of $h_{4}$ with a rising profile might be indicative of radial anisotropy, which can be created by accretion events (e.g \citealt{Hilz2012}; \citealt{Amorisco2017}). This means that the accretion of satellites may have also played an important role in the second phase of the assembly of NGC\,1407. Again, in the {\tt Magneticum} simulation \citep{Forbes2018}, their NGC\,1407-like galaxy was a mixture of both a major merger event and a series of minor mergers, further supporting our results. We note, however, that a full understanding of the orbital structure requires a complete dynamical modelling, along the lines of \citet{vandenBosch2008}.
 
One final clue for the formation of NGC\,1407 is revealed by the location where the changes in the profiles are seen. For example, the transition from a decreasing to an outwards rising $\sigma$ profile occurs around $\sim$0.3--0.4R$\mathrm{_e}$, and the metallicity gradients also seem to flatten around that region, which also corresponds to the rising part of the $h_{4}$ profile. This region interestingly coincides with the location where the IMF of NGC\,1407 transitions from a very bottom-heavy IMF to a Kroupa-like one (see Figure 17 from \hyperlink{vD+17}{vD+17}). IMF variations within a single galaxy make sense under the `two-phase' formation scenario, where the initial physical conditions at the beginning of each phase were completely different (which has recently been reproduced by simulations; e.g \citealt{Barber2019}). \citet{Vaughan2018} suggested that this change in the IMF (in that case for the giant elliptical NGC\,1399) also coincided with the light profile transitioning from an `inner' S\'ersic component to a component suggested to be attributed to the `accreted' stars \citep{Spavone2017}. 

In fact, \citet{Huang2013} showed that the light profile of massive ETGs can generally be fitted by three physically distinctive components rather than by a S\'ersic function. At small scales there is an innermost compact component ($\leq$ 1\,kpc), a fossil-record of the dissipation processes that occurred very early on. They suggested that combined with an intermediate region ($\leq$3\,kpc), together they account for up to $\sim$40$\%$ of the galaxy's luminosity. In the case of high-luminosity ETGs such as NGC\,1407, this region highly resembles the high redshift `red nuggets', which are thought to be the result of the first, dissipative phase (e.g. \citealt{Damjanov2009}; \citealt{Glazebrook2009}). They have typical sizes of $\sim$2\,kpc and stellar masses of $\sim$10$^{11}$M$_\odot$ (e.g. \citealt{Trujillo2007}; \citealt{Buitrago2008}; \citealt{Carrasco2010}). These compact sizes have been further confirmed by the few untouched massive relics of such `red nuggets' found in the nearby Universe (e.g. \citealt{Trujillo2014}; \citealt{Ferre-Mateu2017}; \citealt{Buitrago2018}). In particular, \citet{Huang2013} found that such transition regions occur in NGC\,1407 at $\sim$0.16R$\mathrm{_e}$ and $\sim$0.36R$\mathrm{_e}$, containing 23$\%$ of the light in the galaxy. This coincides with our transition region ($\sim$0.3--0.4R$\mathrm{_e}$), which corresponds to a physical size of $\sim$\,2--3\,kpc and that contains about 20$\%$ of the stellar mass of the galaxy (M$_{\ast}\sim$ 8$\times$10$^{10}$M$_\odot$). Adding all this to the above results, \textit{we therefore suggest that we might be witnessing the transition in NGC\,1407 from the `in-situ' region, corresponding to the `red nugget', into the accreted region of this massive galaxy directly through its galaxy kinematics and stellar populations features}.

\section{Conclusions}

Using two short exposure Keck/KCWI pointings, located centrally and at $\sim$1~R$\mathrm{_e}$, we find NGC\,1407 to have the kinematical features typical of a massive early-type galaxy, i.e., a slow rotator with an inner kinematically distinct core, a centrally peaked velocity dispersion, and high values of anisotropy. Our stellar population analysis is also consistent with that of massive ETGs, i.e., a very old age with short formation timescales of less than 3\,Gyr. The central parts of the galaxy show slightly younger ages that would indicate the effect of a gas-rich major merger at early epochs. These younger ages coincide with the region covered by the small KDC of NGC\,1407. This massive ETG also shows the characteristic strong negative metallicity gradient expected under the `two-phase' scenario.

Overall, our stellar populations are thus all consistent with those reported in the literature and are compatible with a recently simulated mock-NGC\,1407 galaxy. We have also investigated the claims of some outer locations of having anomalously high inferred metallicities. This effect is hard to explain within the `two-phase' scenario. Our 2D metallicity map in this region shows a smooth metallicity gradient with no evidence for the anomalous peaks.

Putting together the kinematic and stellar population results, we suggest that we are probing the transition region from the `in-situ' phase (equivalent to the `red nugget' formed during the first phase) to the accretion-dominated one. The change from a bottom-heavy to a normal IMF, the rise in velocity dispersion, and the increasing anisotropy, all coincide at around 0.3--0.4R$\mathrm{_e}$ (2--3\,kpc).  

This pilot study has shown that KCWI can efficiently obtain accurate kinematic and stellar population properties of ETGs in their central (600s integration, surface brightness $\sim 19$ mag per sq. arcsec) and outer ($4 \times 600$s integration, $\sim 1 R_e$, surface brightness $\sim 22$ mag per sq. arcsec) regions, probing the transition from central `in-situ' dominated formation to accretion dominated. From our investigation, the sensitivity and accuracy of KCWI for such observations appears very comparable to those of the popular MUSE spectrograph on the VLT. We therefore expect that larger surveys of ETGs, and in particular their low surface brightness outer halos, will soon become available using this facility, probing the assembly histories of the most massive galaxies in the Universe.

\bigskip
\section*{Acknowledgments}

AFM has received financial support through the Postdoctoral Junior Leader Fellowship Programme from `La Caixa' Banking Foundation (LCF/BQ/LI18/11630007). AFM, DAF and RMcD also thank the ARC for financial support via DP160101608. RMcD is the recipient of an Australian Research Council Future Fellowship (project number FT150100333). AJR was supported by NSF grant AST-1616710 and as a Research Corporation for Science Advancement Cottrell Scholar. JPB was supported by NSF grant AST-1616598. 

The data presented herein were obtained at the W. M. Keck Observatory, which is operated as a scientific partnership among the California Institute of Technology, the University of California, and the National Aeronautics and Space Administration. The Observatory was made possible by the generous financial support of the W. M. Keck Foundation. The authors wish to recognise and acknowledge the very significant cultural role and reverence that the summit of Maunakea has always had within the indigenous Hawaiian community.  We are most fortunate to have the opportunity to conduct observations from this mountain. \textit{M\=alama ka '\=aina}.

\bibliography{ngc1407}

\begin{thebibliography}{}
\makeatletter
\relax
\def\mn@urlcharsother{\let\do\@makeother \do\$\do\&\do\#\do\^\do\_\do\%\do\~}
\def\mn@doi{\begingroup\mn@urlcharsother \@ifnextchar [ {\mn@doi@}
  {\mn@doi@[]}}
\def\mn@doi@[#1]#2{\def\@tempa{#1}\ifx\@tempa\@empty \href
  {http://dx.doi.org/#2} {doi:#2}\else \href {http://dx.doi.org/#2} {#1}\fi
  \endgroup}
\def\mn@eprint#1#2{\mn@eprint@#1:#2::\@nil}
\def\mn@eprint@arXiv#1{\href {http://arxiv.org/abs/#1} {{\tt arXiv:#1}}}
\def\mn@eprint@dblp#1{\href {http://dblp.uni-trier.de/rec/bibtex/#1.xml}
  {dblp:#1}}
\def\mn@eprint@#1:#2:#3:#4\@nil{\def\@tempa {#1}\def\@tempb {#2}\def\@tempc
  {#3}\ifx \@tempc \@empty \let \@tempc \@tempb \let \@tempb \@tempa \fi \ifx
  \@tempb \@empty \def\@tempb {arXiv}\fi \@ifundefined
  {mn@eprint@\@tempb}{\@tempb:\@tempc}{\expandafter \expandafter \csname
  mn@eprint@\@tempb\endcsname \expandafter{\@tempc}}}

\bibitem[\protect\citeauthoryear{{Amorisco}}{{Amorisco}}{2017}]{Amorisco2017}
{Amorisco} N.~C.,  2017, \mn@doi [\mnras] {10.1093/mnrasl/slx044}, \href
  {http://adsabs.harvard.edu/abs/2017MNRAS.469L..48A} {469, L48}

\bibitem[\protect\citeauthoryear{{Arnold} et~al.,}{{Arnold}
  et~al.}{2014}]{Arnold2014}
{Arnold} J.~A.,  et~al., 2014, \mn@doi [\apj] {10.1088/0004-637X/791/2/80},
  \href {http://adsabs.harvard.edu/abs/2014ApJ...791...80A} {791, 80}

\bibitem[\protect\citeauthoryear{{Bacon} et~al.,}{{Bacon}
  et~al.}{2001}]{Bacon2001}
{Bacon} R.,  et~al., 2001, \mn@doi [\mnras] {10.1046/j.1365-8711.2001.04612.x},
  \href {http://adsabs.harvard.edu/abs/2001MNRAS.326...23B} {326, 23}

\bibitem[\protect\citeauthoryear{{Bacon} et~al.,}{{Bacon}
  et~al.}{2010}]{Bacon2010}
{Bacon} R.,  et~al., 2010, in Ground-based and Airborne Instrumentation for
  Astronomy III. p. 773508, \mn@doi{10.1117/12.856027}

\bibitem[\protect\citeauthoryear{{Barber}, {Schaye}  \& {Crain}}{{Barber}
  et~al.}{2019}]{Barber2019}
{Barber} C.,  {Schaye} J.,   {Crain} R.~A.,  2019, \mn@doi [\mnras]
  {10.1093/mnras/sty3011}, \href
  {http://adsabs.harvard.edu/abs/2019MNRAS.483..985B} {483, 985}

\bibitem[\protect\citeauthoryear{{Beasley}, {Trujillo}, {Leaman}  \&
  {Montes}}{{Beasley} et~al.}{2018}]{Beasley2018}
{Beasley} M.~A.,  {Trujillo} I.,  {Leaman} R.,   {Montes} M.,  2018, \mn@doi
  [\nat] {10.1038/nature25756}, \href
  {http://adsabs.harvard.edu/abs/2018Natur.555..483B} {555, 483}

\bibitem[\protect\citeauthoryear{{Bois} et~al.,}{{Bois}
  et~al.}{2011}]{Bois2011}
{Bois} M.,  et~al., 2011, \mn@doi [\mnras] {10.1111/j.1365-2966.2011.19113.x},
  \href {http://adsabs.harvard.edu/abs/2011MNRAS.416.1654B} {416, 1654}

\bibitem[\protect\citeauthoryear{{Brodie} \& {Strader}}{{Brodie} \&
  {Strader}}{2006}]{Brodie2006}
{Brodie} J.~P.,  {Strader} J.,  2006, \mn@doi [\araa]
  {10.1146/annurev.astro.44.051905.092441}, \href
  {http://adsabs.harvard.edu/abs/2006ARA%26A..44..193B} {44, 193}

\bibitem[\protect\citeauthoryear{{Brodie} et~al.,}{{Brodie}
  et~al.}{2014}]{Brodie2014}
{Brodie} J.~P.,  et~al., 2014, \mn@doi [\apj] {10.1088/0004-637X/796/1/52},
  \href {http://adsabs.harvard.edu/abs/2014ApJ...796...52B} {796, 52}

\bibitem[\protect\citeauthoryear{{Buitrago}, {Trujillo}, {Conselice},
  {Bouwens}, {Dickinson}  \& {Yan}}{{Buitrago} et~al.}{2008}]{Buitrago2008}
{Buitrago} F.,  {Trujillo} I.,  {Conselice} C.~J.,  {Bouwens} R.~J.,
  {Dickinson} M.,   {Yan} H.,  2008, \mn@doi [\apjl] {10.1086/592836}, \href
  {http://adsabs.harvard.edu/abs/2008ApJ...687L..61B} {687, L61}

\bibitem[\protect\citeauthoryear{{Buitrago} et~al.,}{{Buitrago}
  et~al.}{2018}]{Buitrago2018}
{Buitrago} F.,  et~al., 2018, \mn@doi [\aap] {10.1051/0004-6361/201833785},
  \href {http://adsabs.harvard.edu/abs/2018A%26A...619A.137B} {619, A137}

\bibitem[\protect\citeauthoryear{{Bundy} et~al.,}{{Bundy}
  et~al.}{2015}]{Bundy2015}
{Bundy} K.,  et~al., 2015, \mn@doi [\apj] {10.1088/0004-637X/798/1/7}, \href
  {http://adsabs.harvard.edu/abs/2015ApJ...798....7B} {798, 7}

\bibitem[\protect\citeauthoryear{{Cappellari}}{{Cappellari}}{2017}]{Cappellari2017}
{Cappellari} M.,  2017, \mn@doi [\mnras] {10.1093/mnras/stw3020}, \href
  {http://adsabs.harvard.edu/abs/2017MNRAS.466..798C} {466, 798}

\bibitem[\protect\citeauthoryear{{Cappellari} \& {Copin}}{{Cappellari} \&
  {Copin}}{2003}]{Cappellari2003}
{Cappellari} M.,  {Copin} Y.,  2003, \mn@doi [\mnras]
  {10.1046/j.1365-8711.2003.06541.x}, \href
  {http://adsabs.harvard.edu/abs/2003MNRAS.342..345C} {342, 345}

\bibitem[\protect\citeauthoryear{{Cappellari} \& {Emsellem}}{{Cappellari} \&
  {Emsellem}}{2004}]{Cappellari2004}
{Cappellari} M.,  {Emsellem} E.,  2004, \mn@doi [\pasp] {10.1086/381875}, \href
  {http://adsabs.harvard.edu/abs/2004PASP..116..138C} {116, 138}

\bibitem[\protect\citeauthoryear{{Cappellari} et~al.,}{{Cappellari}
  et~al.}{2011}]{Cappellari2011}
{Cappellari} M.,  et~al., 2011, \mn@doi [\mnras]
  {10.1111/j.1365-2966.2010.18174.x}, \href
  {http://adsabs.harvard.edu/abs/2011MNRAS.413..813C} {413, 813}

\bibitem[\protect\citeauthoryear{{Carrasco}, {Conselice}  \&
  {Trujillo}}{{Carrasco} et~al.}{2010}]{Carrasco2010}
{Carrasco} E.~R.,  {Conselice} C.~J.,   {Trujillo} I.,  2010, \mn@doi [\mnras]
  {10.1111/j.1365-2966.2010.16645.x}, \href
  {http://adsabs.harvard.edu/abs/2010MNRAS.405.2253C} {405, 2253}

\bibitem[\protect\citeauthoryear{{Conroy}, {van Dokkum}  \&
  {Villaume}}{{Conroy} et~al.}{2017}]{Conroy2017}
{Conroy} C.,  {van Dokkum} P.~G.,   {Villaume} A.,  2017, \mn@doi [\apj]
  {10.3847/1538-4357/aa6190}, \href
  {http://adsabs.harvard.edu/abs/2017ApJ...837..166C} {837, 166}

\bibitem[\protect\citeauthoryear{{Croom} et~al.,}{{Croom}
  et~al.}{2012}]{Croom2012}
{Croom} S.~M.,  et~al., 2012, \mn@doi [\mnras]
  {10.1111/j.1365-2966.2011.20365.x}, \href
  {http://adsabs.harvard.edu/abs/2012MNRAS.421..872C} {421, 872}

\bibitem[\protect\citeauthoryear{{Daddi} et~al.,}{{Daddi}
  et~al.}{2005}]{Daddi2005}
{Daddi} E.,  et~al., 2005, \mn@doi [\apj] {10.1086/430104}, \href
  {http://adsabs.harvard.edu/abs/2005ApJ...626..680D} {626, 680}

\bibitem[\protect\citeauthoryear{{Damjanov} et~al.,}{{Damjanov}
  et~al.}{2009}]{Damjanov2009}
{Damjanov} I.,  et~al., 2009, \mn@doi [\apj] {10.1088/0004-637X/695/1/101},
  \href {http://adsabs.harvard.edu/abs/2009ApJ...695..101D} {695, 101}

\bibitem[\protect\citeauthoryear{{Dekel}, {Stoehr}, {Mamon}, {Cox}, {Novak}  \&
  {Primack}}{{Dekel} et~al.}{2005}]{Dekel2005}
{Dekel} A.,  {Stoehr} F.,  {Mamon} G.~A.,  {Cox} T.~J.,  {Novak} G.~S.,
  {Primack} J.~R.,  2005, \mn@doi [\nat] {10.1038/nature03970}, \href
  {http://adsabs.harvard.edu/abs/2005Natur.437..707D} {437, 707}

\bibitem[\protect\citeauthoryear{{Ferr{\'e}-Mateu}, {Vazdekis}  \& {de la
  Rosa}}{{Ferr{\'e}-Mateu} et~al.}{2013}]{Ferre-Mateu2013}
{Ferr{\'e}-Mateu} A.,  {Vazdekis} A.,   {de la Rosa} I.~G.,  2013, \mn@doi
  [\mnras] {10.1093/mnras/stt193}, \href
  {http://adsabs.harvard.edu/abs/2013MNRAS.431..440F} {431, 440}

\bibitem[\protect\citeauthoryear{{Ferr{\'e}-Mateu}, {S{\'a}nchez-Bl{\'a}zquez},
  {Vazdekis}  \& {de la Rosa}}{{Ferr{\'e}-Mateu}
  et~al.}{2014}]{Ferre-Mateu2014}
{Ferr{\'e}-Mateu} A.,  {S{\'a}nchez-Bl{\'a}zquez} P.,  {Vazdekis} A.,   {de la
  Rosa} I.~G.,  2014, \mn@doi [\apj] {10.1088/0004-637X/797/2/136}, \href
  {http://adsabs.harvard.edu/abs/2014ApJ...797..136F} {797, 136}

\bibitem[\protect\citeauthoryear{{Ferr{\'e}-Mateu}, {Trujillo},
  {Mart{\'{\i}}n-Navarro}, {Vazdekis}, {Mezcua}, {Balcells}  \&
  {Dom{\'{\i}}nguez}}{{Ferr{\'e}-Mateu} et~al.}{2017}]{Ferre-Mateu2017}
{Ferr{\'e}-Mateu} A.,  {Trujillo} I.,  {Mart{\'{\i}}n-Navarro} I.,  {Vazdekis}
  A.,  {Mezcua} M.,  {Balcells} M.,   {Dom{\'{\i}}nguez} L.,  2017, \mn@doi
  [\mnras] {10.1093/mnras/stx171}, \href
  {http://adsabs.harvard.edu/abs/2017MNRAS.467.1929F} {467, 1929}

\bibitem[\protect\citeauthoryear{{Forbes} \& {Remus}}{{Forbes} \&
  {Remus}}{2018}]{Forbes2018}
{Forbes} D.~A.,  {Remus} R.-S.,  2018, \mn@doi [\mnras]
  {10.1093/mnras/sty1767}, \href
  {http://adsabs.harvard.edu/abs/2018MNRAS.479.4760F} {479, 4760}

\bibitem[\protect\citeauthoryear{{Forbes}, {Sinpetru}, {Savorgnan},
  {Romanowsky}, {Usher}  \& {Brodie}}{{Forbes} et~al.}{2017}]{Forbes2017}
{Forbes} D.~A.,  {Sinpetru} L.,  {Savorgnan} G.,  {Romanowsky} A.~J.,  {Usher}
  C.,   {Brodie} J.,  2017, \mn@doi [\mnras] {10.1093/mnras/stw2604}, \href
  {http://adsabs.harvard.edu/abs/2017MNRAS.464.4611F} {464, 4611}

\bibitem[\protect\citeauthoryear{{Foster} et~al.,}{{Foster}
  et~al.}{2016}]{Foster2016}
{Foster} C.,  et~al., 2016, \mn@doi [\mnras] {10.1093/mnras/stv2947}, \href
  {http://adsabs.harvard.edu/abs/2016MNRAS.457..147F} {457, 147}

\bibitem[\protect\citeauthoryear{{Gerhard}, {Jeske}, {Saglia}  \&
  {Bender}}{{Gerhard} et~al.}{1998}]{Gerhard1998}
{Gerhard} O.,  {Jeske} G.,  {Saglia} R.~P.,   {Bender} R.,  1998, \mn@doi
  [\mnras] {10.1046/j.1365-8711.1998.29511341.x}, \href
  {http://adsabs.harvard.edu/abs/1998MNRAS.295..197G} {295, 197}

\bibitem[\protect\citeauthoryear{{Glazebrook}}{{Glazebrook}}{2009}]{Glazebrook2009}
{Glazebrook} K.,  2009, \mn@doi [\nat] {10.1038/460694a}, \href
  {http://adsabs.harvard.edu/abs/2009Natur.460..694G} {460, 694}

\bibitem[\protect\citeauthoryear{{Gould}}{{Gould}}{1993}]{Gould1993}
{Gould} A.,  1993, \mn@doi [\apj] {10.1086/172181}, \href
  {http://adsabs.harvard.edu/abs/1993ApJ...403...37G} {403, 37}

\bibitem[\protect\citeauthoryear{{Greene} et~al.,}{{Greene}
  et~al.}{2019}]{Greene2019}
{Greene} J.~E.,  et~al., 2019, arXiv e-prints, \href
  {http://adsabs.harvard.edu/abs/2019arXiv190101271G} {}

\bibitem[\protect\citeauthoryear{{Hernquist} \& {Barnes}}{{Hernquist} \&
  {Barnes}}{1991}]{Hernquist1991}
{Hernquist} L.,  {Barnes} J.~E.,  1991, \mn@doi [\nat] {10.1038/354210a0},
  \href {http://adsabs.harvard.edu/abs/1991Natur.354..210H} {354, 210}

\bibitem[\protect\citeauthoryear{{Hilz}, {Naab}, {Ostriker}, {Thomas},
  {Burkert}  \& {Jesseit}}{{Hilz} et~al.}{2012}]{Hilz2012}
{Hilz} M.,  {Naab} T.,  {Ostriker} J.~P.,  {Thomas} J.,  {Burkert} A.,
  {Jesseit} R.,  2012, \mn@doi [\mnras] {10.1111/j.1365-2966.2012.21541.x},
  \href {http://adsabs.harvard.edu/abs/2012MNRAS.425.3119H} {425, 3119}

\bibitem[\protect\citeauthoryear{{Hilz}, {Naab}  \& {Ostriker}}{{Hilz}
  et~al.}{2013}]{Hilz2013}
{Hilz} M.,  {Naab} T.,   {Ostriker} J.~P.,  2013, \mn@doi [\mnras]
  {10.1093/mnras/sts501}, \href
  {http://adsabs.harvard.edu/abs/2013MNRAS.429.2924H} {429, 2924}

\bibitem[\protect\citeauthoryear{{Holley-Bockelmann} \&
  {Richstone}}{{Holley-Bockelmann} \&
  {Richstone}}{2000}]{Holley-Bockelmann2000}
{Holley-Bockelmann} K.,  {Richstone} D.~O.,  2000, \mn@doi [\apj]
  {10.1086/308447}, \href {http://adsabs.harvard.edu/abs/2000ApJ...531..232H}
  {531, 232}

\bibitem[\protect\citeauthoryear{{Jesseit}, {Naab}, {Peletier}  \&
  {Burkert}}{{Jesseit} et~al.}{2007}]{Jesseit2007}
{Jesseit} R.,  {Naab} T.,  {Peletier} R.~F.,   {Burkert} A.,  2007, \mn@doi
  [\mnras] {10.1111/j.1365-2966.2007.11524.x}, \href
  {http://adsabs.harvard.edu/abs/2007MNRAS.376..997J} {376, 997}

\bibitem[\protect\citeauthoryear{{Johnston}, {Hau}, {Coccato}  \&
  {Herrera}}{{Johnston} et~al.}{2018}]{Johnston2018}
{Johnston} E.~J.,  {Hau} G.~K.~T.,  {Coccato} L.,   {Herrera} C.,  2018,
  \mn@doi [\mnras] {10.1093/mnras/sty2048}, \href
  {http://adsabs.harvard.edu/abs/2018MNRAS.480.3215J} {480, 3215}

\bibitem[\protect\citeauthoryear{{Ma}, {Greene}, {McConnell}, {Janish},
  {Blakeslee}, {Thomas}  \& {Murphy}}{{Ma} et~al.}{2014}]{Ma2014}
{Ma} C.-P.,  {Greene} J.~E.,  {McConnell} N.,  {Janish} R.,  {Blakeslee} J.~P.,
   {Thomas} J.,   {Murphy} J.~D.,  2014, \mn@doi [\apj]
  {10.1088/0004-637X/795/2/158}, \href
  {http://adsabs.harvard.edu/abs/2014ApJ...795..158M} {795, 158}

\bibitem[\protect\citeauthoryear{{Mart{\'{\i}}n-Navarro}, {La Barbera},
  {Vazdekis}, {Falc{\'o}n-Barroso}  \& {Ferreras}}{{Mart{\'{\i}}n-Navarro}
  et~al.}{2015}]{Martin-Navarro2015}
{Mart{\'{\i}}n-Navarro} I.,  {La Barbera} F.,  {Vazdekis} A.,
  {Falc{\'o}n-Barroso} J.,   {Ferreras} I.,  2015, \mn@doi [\mnras]
  {10.1093/mnras/stu2480}, \href
  {http://adsabs.harvard.edu/abs/2015MNRAS.447.1033M} {447, 1033}

\bibitem[\protect\citeauthoryear{{McDermid} et~al.,}{{McDermid}
  et~al.}{2006}]{McDermid2006}
{McDermid} R.~M.,  et~al., 2006, \mn@doi [\mnras]
  {10.1111/j.1365-2966.2006.11065.x}, \href
  {http://adsabs.harvard.edu/abs/2006MNRAS.373..906M} {373, 906}

\bibitem[\protect\citeauthoryear{{McDermid} et~al.,}{{McDermid}
  et~al.}{2015}]{McDermid2015}
{McDermid} R.~M.,  et~al., 2015, \mn@doi [\mnras] {10.1093/mnras/stv105}, \href
  {http://adsabs.harvard.edu/abs/2015MNRAS.448.3484M} {448, 3484}

\bibitem[\protect\citeauthoryear{{Morrissey} et~al.,}{{Morrissey}
  et~al.}{2018}]{Morrissey2018}
{Morrissey} P.,  et~al., 2018, \mn@doi [\apj] {10.3847/1538-4357/aad597}, \href
  {http://adsabs.harvard.edu/abs/2018ApJ...864...93M} {864, 93}

\bibitem[\protect\citeauthoryear{{Naab}, {Johansson}  \& {Ostriker}}{{Naab}
  et~al.}{2009}]{Naab2009}
{Naab} T.,  {Johansson} P.~H.,   {Ostriker} J.~P.,  2009, \mn@doi [\apj]
  {10.1088/0004-637X/699/2/L178}, \href
  {http://adsabs.harvard.edu/abs/2009ApJ...699L.178N} {699, L178}

\bibitem[\protect\citeauthoryear{{Oser}, {Ostriker}, {Naab}, {Johansson}  \&
  {Burkert}}{{Oser} et~al.}{2010}]{Oser2010}
{Oser} L.,  {Ostriker} J.~P.,  {Naab} T.,  {Johansson} P.~H.,   {Burkert} A.,
  2010, \mn@doi [\apj] {10.1088/0004-637X/725/2/2312}, \href
  {http://adsabs.harvard.edu/abs/2010ApJ...725.2312O} {725, 2312}

\bibitem[\protect\citeauthoryear{{Pastorello}, {Forbes}, {Foster}, {Brodie},
  {Usher}, {Romanowsky}, {Strader}  \& {Arnold}}{{Pastorello}
  et~al.}{2014}]{Pastorello2014}
{Pastorello} N.,  {Forbes} D.~A.,  {Foster} C.,  {Brodie} J.~P.,  {Usher} C.,
  {Romanowsky} A.~J.,  {Strader} J.,   {Arnold} J.~A.,  2014, \mn@doi [\mnras]
  {10.1093/mnras/stu937}, \href
  {http://adsabs.harvard.edu/abs/2014MNRAS.442.1003P} {442, 1003}

\bibitem[\protect\citeauthoryear{{Peng} et~al.,}{{Peng}
  et~al.}{2006}]{Peng2006}
{Peng} E.~W.,  et~al., 2006, \mn@doi [\apj] {10.1086/498210}, \href
  {http://adsabs.harvard.edu/abs/2006ApJ...639...95P} {639, 95}

\bibitem[\protect\citeauthoryear{{Pota} et~al.,}{{Pota}
  et~al.}{2015}]{Pota2015}
{Pota} V.,  et~al., 2015, \mn@doi [\mnras] {10.1093/mnras/stv831}, \href
  {http://adsabs.harvard.edu/abs/2015MNRAS.450.3345P} {450, 3345}

\bibitem[\protect\citeauthoryear{{Proctor}, {Forbes}, {Romanowsky}, {Brodie},
  {Strader}, {Spolaor}, {Mendel}  \& {Spitler}}{{Proctor}
  et~al.}{2009}]{Proctor2009}
{Proctor} R.~N.,  {Forbes} D.~A.,  {Romanowsky} A.~J.,  {Brodie} J.~P.,
  {Strader} J.,  {Spolaor} M.,  {Mendel} J.~T.,   {Spitler} L.,  2009, \mn@doi
  [\mnras] {10.1111/j.1365-2966.2009.15137.x}, \href
  {http://adsabs.harvard.edu/abs/2009MNRAS.398...91P} {398, 91}

\bibitem[\protect\citeauthoryear{{Remus}, {Dolag}, {Naab}, {Burkert},
  {Hirschmann}, {Hoffmann}  \& {Johansson}}{{Remus} et~al.}{2017}]{Remus2017}
{Remus} R.-S.,  {Dolag} K.,  {Naab} T.,  {Burkert} A.,  {Hirschmann} M.,
  {Hoffmann} T.~L.,   {Johansson} P.~H.,  2017, \mn@doi [\mnras]
  {10.1093/mnras/stw2594}, \href
  {http://adsabs.harvard.edu/abs/2017MNRAS.464.3742R} {464, 3742}

\bibitem[\protect\citeauthoryear{{Rodriguez-Gomez} et~al.,}{{Rodriguez-Gomez}
  et~al.}{2016}]{RodriguezGomez2016}
{Rodriguez-Gomez} V.,  et~al., 2016, \mn@doi [\mnras] {10.1093/mnras/stw456},
  \href {http://adsabs.harvard.edu/abs/2016MNRAS.458.2371R} {458, 2371}

\bibitem[\protect\citeauthoryear{{Rusli} et~al.,}{{Rusli}
  et~al.}{2013}]{Rusli2013}
{Rusli} S.~P.,  et~al., 2013, \mn@doi [\aj] {10.1088/0004-6256/146/3/45}, \href
  {http://adsabs.harvard.edu/abs/2013AJ....146...45R} {146, 45}

\bibitem[\protect\citeauthoryear{{S{\'a}nchez} et~al.,}{{S{\'a}nchez}
  et~al.}{2012}]{Sanchez2012}
{S{\'a}nchez} S.~F.,  et~al., 2012, \mn@doi [\aap]
  {10.1051/0004-6361/201117353}, \href
  {http://adsabs.harvard.edu/abs/2012A%26A...538A...8S} {538, A8}

\bibitem[\protect\citeauthoryear{{Spavone} et~al.,}{{Spavone}
  et~al.}{2017}]{Spavone2017}
{Spavone} M.,  et~al., 2017, \mn@doi [\aap] {10.1051/0004-6361/201629111},
  \href {http://adsabs.harvard.edu/abs/2017A%26A...603A..38S} {603, A38}

\bibitem[\protect\citeauthoryear{{Spolaor}, {Forbes}, {Hau}, {Proctor}  \&
  {Brough}}{{Spolaor} et~al.}{2008a}]{Spolaor2008}
{Spolaor} M.,  {Forbes} D.~A.,  {Hau} G.~K.~T.,  {Proctor} R.~N.,   {Brough}
  S.,  2008a, \mn@doi [\mnras] {10.1111/j.1365-2966.2008.12891.x}, \href
  {http://adsabs.harvard.edu/abs/2008MNRAS.385..667S} {385, 667}

\bibitem[\protect\citeauthoryear{{Spolaor}, {Forbes}, {Proctor}, {Hau}  \&
  {Brough}}{{Spolaor} et~al.}{2008b}]{Spolaor2008a}
{Spolaor} M.,  {Forbes} D.~A.,  {Proctor} R.~N.,  {Hau} G.~K.~T.,   {Brough}
  S.,  2008b, \mn@doi [\mnras] {10.1111/j.1365-2966.2008.12892.x}, \href
  {http://adsabs.harvard.edu/abs/2008MNRAS.385..675S} {385, 675}

\bibitem[\protect\citeauthoryear{{Thomas}, {Maraston}, {Bender}  \& {Mendes de
  Oliveira}}{{Thomas} et~al.}{2005}]{Thomas2005}
{Thomas} D.,  {Maraston} C.,  {Bender} R.,   {Mendes de Oliveira} C.,  2005,
  \mn@doi [\apj] {10.1086/426932}, \href
  {http://adsabs.harvard.edu/abs/2005ApJ...621..673T} {621, 673}

\bibitem[\protect\citeauthoryear{{Thomas}, {Saglia}, {Bender}, {Thomas},
  {Gebhardt}, {Magorrian}, {Corsini}  \& {Wegner}}{{Thomas}
  et~al.}{2007}]{Thomas2007}
{Thomas} J.,  {Saglia} R.~P.,  {Bender} R.,  {Thomas} D.,  {Gebhardt} K.,
  {Magorrian} J.,  {Corsini} E.~M.,   {Wegner} G.,  2007, \mn@doi [\mnras]
  {10.1111/j.1365-2966.2007.12434.x}, \href
  {http://adsabs.harvard.edu/abs/2007MNRAS.382..657T} {382, 657}

\bibitem[\protect\citeauthoryear{{Trujillo}, {Conselice}, {Bundy}, {Cooper},
  {Eisenhardt}  \& {Ellis}}{{Trujillo} et~al.}{2007}]{Trujillo2007}
{Trujillo} I.,  {Conselice} C.~J.,  {Bundy} K.,  {Cooper} M.~C.,  {Eisenhardt}
  P.,   {Ellis} R.~S.,  2007, \mnras, 382, 109

\bibitem[\protect\citeauthoryear{{Trujillo}, {Ferr{\'e}-Mateu}, {Balcells},
  {Vazdekis}  \& {S{\'a}nchez-Bl{\'a}zquez}}{{Trujillo}
  et~al.}{2014}]{Trujillo2014}
{Trujillo} I.,  {Ferr{\'e}-Mateu} A.,  {Balcells} M.,  {Vazdekis} A.,
  {S{\'a}nchez-Bl{\'a}zquez} P.,  2014, \mn@doi [\apjl]
  {10.1088/2041-8205/780/2/L20}, \href
  {http://adsabs.harvard.edu/abs/2014ApJ...780L..20T} {780, L20}

\bibitem[\protect\citeauthoryear{{Vaughan}, {Davies}, {Zieleniewski}  \&
  {Houghton}}{{Vaughan} et~al.}{2018}]{Vaughan2018}
{Vaughan} S.~P.,  {Davies} R.~L.,  {Zieleniewski} S.,   {Houghton} R.~C.~W.,
  2018, \mn@doi [\mnras] {10.1093/mnras/sty1434}, \href
  {http://adsabs.harvard.edu/abs/2018MNRAS.479.2443V} {479, 2443}

\bibitem[\protect\citeauthoryear{{Vazdekis}, {Koleva}, {Ricciardelli},
  {R{\"o}ck}  \& {Falc{\'o}n-Barroso}}{{Vazdekis} et~al.}{2016}]{Vazdekis2016}
{Vazdekis} A.,  {Koleva} M.,  {Ricciardelli} E.,  {R{\"o}ck} B.,
  {Falc{\'o}n-Barroso} J.,  2016, \mn@doi [\mnras] {10.1093/mnras/stw2231},
  \href {http://adsabs.harvard.edu/abs/2016MNRAS.463.3409V} {463, 3409}

\bibitem[\protect\citeauthoryear{{Veale}, {Ma}, {Greene}, {Thomas},
  {Blakeslee}, {Walsh}  \& {Ito}}{{Veale} et~al.}{2018}]{Veale2018}
{Veale} M.,  {Ma} C.-P.,  {Greene} J.~E.,  {Thomas} J.,  {Blakeslee} J.~P.,
  {Walsh} J.~L.,   {Ito} J.,  2018, \mn@doi [\mnras] {10.1093/mnras/stx2717},
  \href {http://adsabs.harvard.edu/abs/2018MNRAS.473.5446V} {473, 5446}

\bibitem[\protect\citeauthoryear{{Wasserman} et~al.,}{{Wasserman}
  et~al.}{2018}]{Wasserman2018}
{Wasserman} A.,  et~al., 2018, \mn@doi [\apj] {10.3847/1538-4357/aad236}, \href
  {http://adsabs.harvard.edu/abs/2018ApJ...863..130W} {863, 130}

\bibitem[\protect\citeauthoryear{{de La Rosa}, {La Barbera}, {Ferreras}  \& {de
  Carvalho}}{{de La Rosa} et~al.}{2011}]{delaRosa2011}
{de La Rosa} I.~G.,  {La Barbera} F.,  {Ferreras} I.,   {de Carvalho} R.~R.,
  2011, \mn@doi [\mnras] {10.1111/j.1745-3933.2011.01146.x}, \href
  {http://adsabs.harvard.edu/abs/2011MNRAS.418L..74D} {418, L74}

\bibitem[\protect\citeauthoryear{{van Dokkum}, {Conroy}, {Villaume}, {Brodie}
  \& {Romanowsky}}{{van Dokkum} et~al.}{2017}]{vanDokkum2017}
{van Dokkum} P.,  {Conroy} C.,  {Villaume} A.,  {Brodie} J.,   {Romanowsky}
  A.~J.,  2017, \mn@doi [\apj] {10.3847/1538-4357/aa7135}, \href
  {http://adsabs.harvard.edu/abs/2017ApJ...841...68V} {841, 68}

\bibitem[\protect\citeauthoryear{{van den Bosch}, {van de Ven}, {Verolme},
  {Cappellari}  \& {de Zeeuw}}{{van den Bosch} et~al.}{2008}]{vandenBosch2008}
{van den Bosch} R.~C.~E.,  {van de Ven} G.,  {Verolme} E.~K.,  {Cappellari} M.,
    {de Zeeuw} P.~T.,  2008, \mn@doi [\mnras]
  {10.1111/j.1365-2966.2008.12874.x}, \href
  {http://adsabs.harvard.edu/abs/2008MNRAS.385..647V} {385, 647}

\makeatother
\end{thebibliography}
\bibliographystyle{mnras}

\end{document}